# Habitat transition in the evolution of bacteria and archaea

Alexander L. Jaffe[1,2], Cindy J. Castelle[3,4], and Jillian F. Banfield[3-6*]

[1]Department of Plant and Microbial Biology, University of California, Berkeley, CA
[2]Department of Earth System Science, Stanford University, Stanford, CA
[3]Innovative Genomics Institute, University of California, Berkeley, CA
[4]Department of Earth and Planetary Science, University of California, Berkeley, CA
[5]Department of Environmental Science, Policy, and Management, University of California, Berkeley, CA
[6]Chan Zuckerberg Biohub, San Francisco, CA

[*]Corresponding author email: jbanfield@berkeley.edu

## Abstract

Related groups of microbes are widely distributed across Earth's habitats, implying numerous dispersal and adaptation events over evolutionary time. However, to date, relatively little is known about the characteristics and mechanisms of these habitat transitions, particularly for populations that reside in animal microbiomes. Here, we review the existing literature concerning habitat transitions among a variety of bacterial and archaeal lineages, considering the frequency of migration events, potential environmental barriers, and mechanisms of adaptation to new physicochemical conditions, including the modification of protein inventories and other genomic characteristics. Cells dependent on microbial hosts, particularly bacteria from the Candidate Phyla Radiation (CPR), have undergone repeated habitat transitions from environmental sources into animal microbiomes. We compare their trajectories to those of both free-living cells - including the Melainabacteria, Elusimicrobia, and methanogenic archaea - as well cellular endosymbionts and bacteriophages, which have made similar transitions. We conclude by highlighting major related topics that may be worthy of future study.

**Keywords:** habitat transition, microbial evolution, dispersal and adaptation, Candidate Phyla Radiation bacteria, human microbiome evolution, environmental microbiology

## Introduction

Since the origin of life on Earth, the diversification and success of microbes has been intimately intertwined with their physical and chemical environments. Over time, geological change enabled dispersal of existing lineages into new habitats with differing physical conditions and resources. On the other hand, biological innovations dramatically altered the composition of their environments - for example, the development of oxygenic photosynthesis by ancestors of the Cyanobacteria that led to drastic reduction in iron availability (35). Such changes to the atmosphere, oceans, aquifers, and soils then in turn provided new opportunities for life, driving dispersal into the new habitats and ultimately leading to biological innovations such as aerobic respiration. In later Earth history, the evolution of plants and animals would provide even more new habitats into which environmental lineages could disperse and thrive. Presumably, such processes eventually led to the establishment of host-associated microbiomes populated by organisms that are distinct from their environmental relatives (6, 49, 91).

Of course, present-day surveys of microbial community composition across Earth's ecosystems speak only to what currently is, and obscure the complex patterns of diversification, dispersal, extinction, and establishment that likely shaped microbial lineages over time. How, when, and where habitat transitions occurred is still almost entirely unknown for most bacterial and archaeal lineages. This stands in contrast to our understanding of biogeography and evolutionary history for animals, where principles governing the colonization of new habitats like islands, ecological succession, and evolutionary radiation following colonization are better understood (e.g. 53). Recent reviews have established that some of these principles are applicable to the microbial world (24, 80); however, the general constraints on and requirements for adaptation to the innumerable habitats populated by microbes remain little understood.

Fortunately, in the last decade, an ever-growing repository of genomic information has enabled comparative genomic analyses for many lineages with representatives spanning sediment, soil, plant, aquifers, lakes, rivers, oceans, animal, and terrestrial habitats. However, few studies have drawn upon this body of literature to identify commonalities and differences among habitat transition processes that underlie microbial distribution patterns. Here, we synthesize the available literature to discuss the frequency, constraints, and genomic changes that are associated with microbial habitat transitions, which we define as a migration from one niche to a significantly different one (e.g., groundwater to human microbiome, or ocean to freshwater) that leads to lasting genomic change. We also consider the impact of obligate dependency on the ability of organisms and viruses to disperse and ultimately establish in new habitats, focusing on the particular case of the Candidate Phyla Radiation (CPR), which are typically episymbionts of other cells (14, 42, 56). We conclude with an overview of major open questions in the study of habitat transitions and discuss how their study could serve our understanding of the past, present, and future of the human microbiome.

## I. The where, when, and how much of microbial habitat transition

The oft-invoked adage "everything is everywhere, but the environment selects" (27 and references therein) implies that dispersal is facile, but the outcome is modulated by the ability of dispersed cells to take up long-term residence in new habitats. By considering the extant distributions of bacteria and archaea, it is apparent that members of most phylum level lineages are broadly distributed across major habitat types (91). While transitions between some of these habitat types may be 'dead ends' - e.g., the irreversible symbioses developed by some insect symbionts (63), or the endosymbiotic process that led to mitochondria and plastids (30) - beyond these extreme cases, there is likely a continuum of transition frequency among different habitat types. The frequency of successful habitat transitions among lineages is an interesting yet understudied question in microbial ecology, and an important one, as habitat changes are direct drivers of evolution and diversity generation. We know little about how often bacteria or archaea successfully disperse to, and establish within, new environments and how this varies by lineages and habitat type. It is also unclear which types of habitat change are most common.

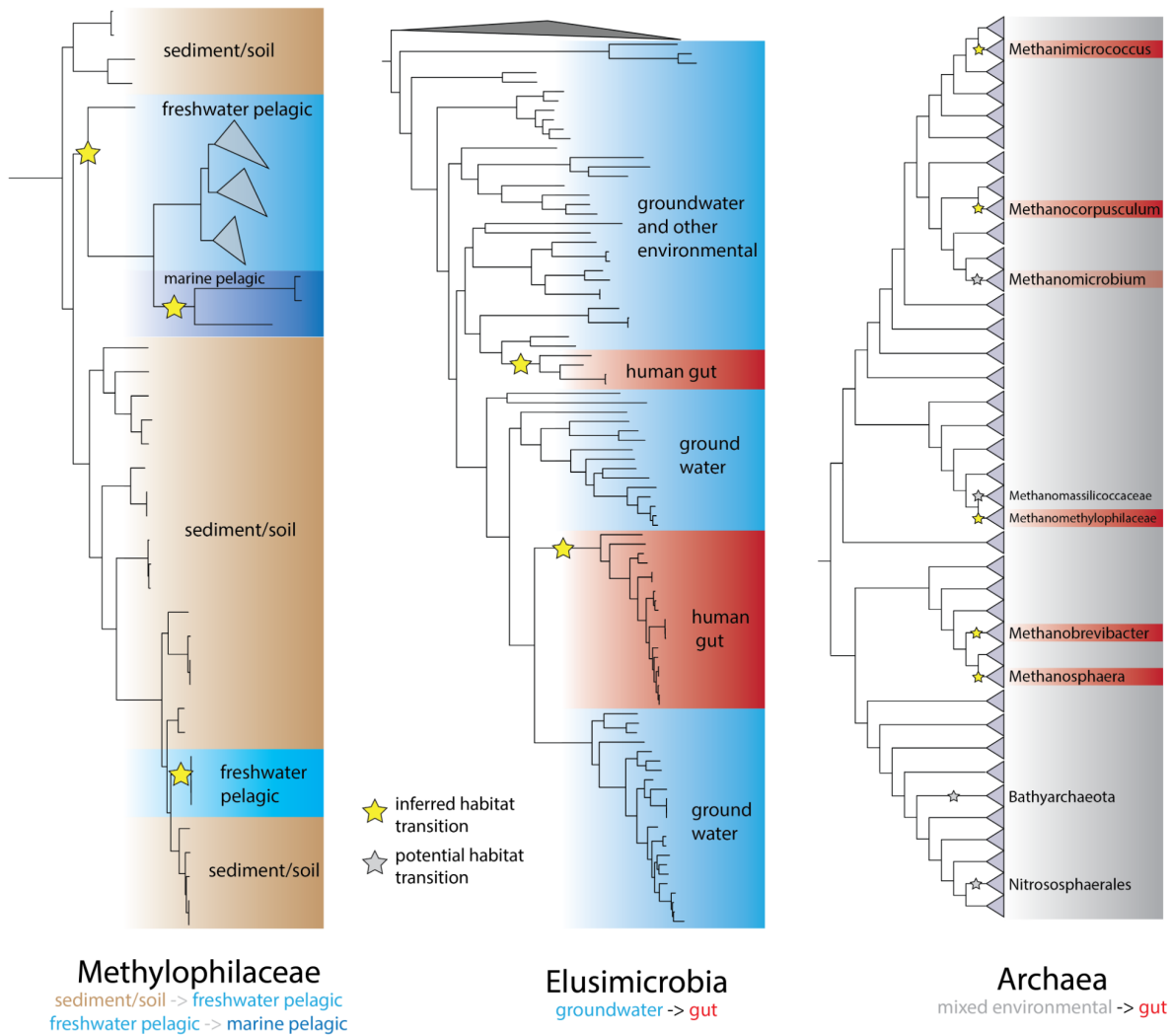

**Figure 1.** Habitat transitions viewed within a phylogenetic framework. Each tree represents a lineage (or set of lineages) shaded with their primary habitat affiliation. The approximate locations of inferred or potential habitat transitions on the topologies are indicated with stars. Trees are adapted from Salcher et al. 2019, Meheust et al. 2020, and Thomas et al. 2022, respectively.

### The diversity of lineages and transitions

Fortunately, molecular phylogenetics provides powerful tools for answering such questions. To study habitat transition in a lineage or set of related lineages, a 'species' tree describing the evolutionary relationships between extant organisms is constructed, onto which environmental information describing the source for those organisms is mapped (Fig. 1). Tree regions where habitat affiliation is consistent in subgroups (or clades) but different in phylogenetically flanking groups are interpreted as evidence for transitions (Fig. 1). By searching for such regions, it is possible to determine not only how many times lineages successfully transitioned from one habitat to another but, in some cases, also the source habitat and thus the directionality of these transitions. Early studies applying phylogenetic frameworks to amplicon sequencing data (such as regions of the 16S rRNA gene) found that more closely related sequences were more likely to share a habitat, suggesting that habitat association is largely conserved and thus that transitions are rare, at least over short evolutionary time scales (96). However, there is ample evidence of transitions over longer timescales (96). Importantly, these analyses are highly dependent on sampling and the incidence of recent transitions may be underestimated if groups are not adequately sampled. This problem is exacerbated if lineages exist at or below detection thresholds (69).

In the last decade, new research has increasingly drawn on collections of draft (and sometimes complete) genomes, many of which were generated via genome-resolved metagenomics (94) and access bacterial and archaeal lineages from a wide variety of environment types. One advantage of genome-centric approaches is that multiple conserved, phylogenetically informative genes can be concatenated and used to build species trees that may more accurately depict evolutionary relationships than single gene approaches (38). Studies utilizing these methods suggest that habitat transition from one natural environment to another is a widespread phenomenon among bacteria, archaea, and eukaryotes (45). Among bacteria, numerous transitions have been observed between sediment/soil and aquatic ecosystems (e.g., in the phyla Planctomycetes and Chloroflexi) (3, 60), as well as between the ocean and terrestrial environments (e.g., among Flavobacteriaceae and Actinobacteria) (70, 99). Also observed are transitions between the warm, sunlit ocean surface and the deeper, colder layers in some Gammaproteobacteria (87) and between marine and freshwater environments in clades like the Alphaproteobacteria (including SAR11) (31, 52, 83), Methylophilaceae (73, 77) (Fig.1a), and some Thaumarchaeota (75).

Phylogenetic reconstructions also show evidence for the movement of bacteria from the environment (often groundwater) into the human/animal oral and gut microbiomes. Examples include the Elusimicrobia (Fig. 1b) (59), certain lineages in the Candidate Phyla Radiation (44, 58, 81), and non-photosynthetic relatives of the Cyanobacteria called the Melainabacteria (28, 57). However, there is variability in the number of times any given lineage established in human/animal microbiomes, including among the CPR bacteria, with some lineages like the Absconditabacteria seemingly colonizing only once (although this group remains highly undersampled). Other better sampled CPR bacteria such as the Saccharibacteria clearly underwent transitions multiple times (44, 58). Similarly, among archaea, phylogenetic data suggest independent transitions of at least five lineages into the human gut from environmental sources, including multiple groups of archaea known to produce methane as well as those that do not (e.g., gut Thaumarchaeota) (89) (Fig. 1c).

**Timescales relevant for habitat transition**

Another important question concerns the timescales over which habitat transitions take place. In principle, habitat changes likely exist on a spectrum ranging from very recent (on the order of hours or days) to extremely ancient (on the order of billions or millions of years). In the well-studied *Vibrio fischeri*-bobtail squid (*Euprymna scolopes*) symbiosis, bacterial symbionts are recruited from surrounding seawater to the animal tissues at the onset of symbiosis (93). Subsequently, at the beginning of each day, approximately 90% of *Vibrio* cells are vented out of the light organ back to their original planktonic niche (67). Thus, in contrast to rarer events observed in other lineages, habitat transitions occur nearly constantly for *V. fischeri*, and, remarkably, are accompanied by apparently reversible changes to cell size and motility structures (76, 93). Similar dynamics are also observable in other horizontally-transmitted symbioses such as those in marine mollusks, annelids, and root nodules of terrestrial plants (10). However, in this review article, we primarily consider those habitat transitions that have happened over long evolutionary timescales, as these transitions can be more easily resolved by phylogenetic approaches.

With the exception of rare cases in which a fossil record exists for insect hosts of symbiotic bacteria (e.g. 62), time-calibrating more ancient habitat transitions is generally not possible. Phylogenetic trees, however, can give clues as to the relative timing of transitions within lineages through branching order. In phylogenetic trees, events that occurred at deeper (more interior) nodes are inferred to be more ancient, whereas those occurring at more shallow (exterior) nodes occurred closer to the present. For example, in the case of Actinobacteria, extant marine lineages are often closely related to (i.e., separated only by short branches from) extant non-marine lineages, suggesting that transitions between marine and freshwater habitats occurred relatively recently (70). Among methanogenic archaea that have migrated from the environment to the gut, phylogenetic data suggest a mix of relatively recent and more ancient transitions, with some occurring early in or prior to the emergence of genera (e.g. *Methanosphaera)*, and others later on (Fig. 1c) (89). Such early transitions may be consistent with the inference that methanogens were present in the microbiome of the last common ancestor of mammals (98).

In general, though, it is important to recognize that current analysis methods are likely to under-recover ancient habitat transitions if some lineages resulting from these transitions survived for some time but ultimately did not persist. Thus, all incidences of habitat transition estimated from contemporary sampling efforts should likely be interpreted as conservative minima.

## II. Barriers to microbial habitat transition

Are some microbial habitat transitions 'more difficult' or 'easier' than others, and if so, why? One relatively simplistic answer is that dispersal is in itself a barrier that can impact the probability of successful habitat transitions (88). Dispersal limitation may be extreme if the initial and destination habitats are separated by a hostile environment (97), and thus, frequencies of transitions may be higher where there is environmental contiguity. This idea is supported by the observation that some bacteria found in mammalian mouths and guts are abundant in freshwater, including drinking water (40), and those in the dolphin mouth most closely related to those from seawater (44).

Another major factor determining whether lineages successfully establish once dispersal has occurred is their ability to meet the challenges posed by new conditions. In particular, a considerable body of research in both the prokaryotic and eukaryotic worlds has examined the 'salt barrier', i.e., the challenges associated with transition between saline marine ecosystems and less saline terrestrial ones, including freshwater (45, 51). For an organism to cross the salinity barrier likely involves complex genetic changes, including but not limited to transporters that use different ions, osmotic stress response mechanisms, lipid composition, and central metabolic and biosynthetic pathways (29, 69, 70). Supporting this, evolutionary analysis shows that multiple lineages of bacteria and archaea have gained sodium transporters, components involved in ectoine synthesis and synthesis of other alternative osmolytes like proline, and the Na+-translocating (NQR) NADH:quinone dehydrogenases complex in adapting to marine environments (Table 1)(1, 29, 70, 74, 83, 99). In contrast, transitions to less saline environments often entail the loss of Na+ based transporters, replacement of the NQR with an $H^+$-translocating equivalent, and certain mechanosensitive channels that mitigate osmotic downshock (Table 1)(83, 99). Finally, some adaptations to higher salinity - namely, shifting amino acids towards higher isoelectric points - affect large portions of the proteome and thus may require longer stretches of time to achieve (13, 84).

The extent of genetic changes needed to cross the salt barrier has been used to explain why microbial communities composition differs greatly along salinity gradients (4, 29, 55) and to suggest that microbial transitions between high and low salinity habitats are rare (51, 52), although this is debated (7, 69). In some cases, more gradual transition could be facilitated via stepwise colonization of habitats along a salinity gradient (e.g., from freshwater, to brackish, and ultimately, to marine sites) (41). It is also important to note that many terrestrial, near-surface ecosystems like soils have microhabitats that become saline as the matrix dries out (18). Thus, soil-associated microorganisms may be primed for colonization of environments with a broad range of salinities.

## III. Genomic changes associated with habitat transition

In recent years, a growing number of studies have directly compared whole genomic information from closely related lineages across environment types, revealing large groups of proteins both of known and unknown function that are associated with various habitat transitions. The identification of those proteins that are currently hypothetical raise the possibility that important adaptive changes are yet to be characterized. However, among those proteins that can be confidently annotated with current methods, common trends are emerging. In particular, numerous transitions have involved genes related to stress response, energy and carbon metabolism, and motility, suggesting that habitat change often exerts similar pressures on microbes, although specific genetic solutions may vary with environment type. These genetic features, grouped by type of habitat transition, are summarized in Table 1.

**Changes in protein coding genes**

Existing genomic evidence suggests that microbes often shift their metabolic strategies in response to changes in the spectrum of available nutrients (31, 99). For example, adaptation to the pelagic or the animal microbiome is often accompanied by loss of central metabolic capacities, including those involved in nitrogen, sulfur, and carbon metabolism, compared to environmental relatives with larger genetic inventories. A particularly striking example is that of the Elusimicrobia, where gut representatives likely lost autotrophic, heterotrophic, and diverse respiratory strategies found in groundwater clades in favor of more simplistic, fermentation-based metabolic platforms (Table 1)(59). However, despite gene loss as a predominant force, gene acquisition is also apparent in lineages undergoing such transitions; for example, genes for the gain of uptake systems for nitrogenous compounds by organisms adapting to the freshwater pelagic environment from soil or sediment (Table 1) (3, 77). Similarly, numerous lineages in both the Archaea and Bacteria have acquired genes encoding rhodopsin - light driven ion pumps that can be coupled to ATP generation - upon adaptation to sunlit aquatic environments, where they contribute to energy production in otherwise oligotrophic conditions (37, 47, 77, 100). Finally, the specific mix of polysaccharides, peptides, and fatty acids present in a given habitat also shapes some organisms' inventory of degradative enzymes and may have impacts on their evolutionary diversification (99, 100).

Oxygen availability can also serve as an important determinant of microbial gene repertoires. For example, evolutionary analysis has suggested that early Thaumarchaeota transitioning from anaerobic to aerobic terrestrial environments lost the Wood-Ljungdahl pathway and gained a key pathway enabling energy production from ammonia (Table 1) (74, 82). The acquisition or loss of some genes involved in oxidative stress response has also been observed, a notable example being the cytochrome o ubiquinol oxidase operon found in some Saccharibacteria (CPR). These genes are thought to function in detoxification or rudimentary oxygen metabolism in organisms that are generally obligate fermenters (46, 66, 85) and are conserved among environmental lineages, including those in soil. The operon is absent in closely related lineages from anoxic oral sites, suggesting that it was lost during the process of transition from the environment (44). However, it is also possible that the capacity to cope with oxygen involved more recent gene acquisition. Finally, as discussed above, adaptation to osmotic stress - particularly among lineages crossing the 'salt barrier' - has been a major driver of proteomic change over time.

**Trends in genome size**

The loss of protein coding genes underlies the broader pattern of genome reduction that is commonly observed during the course of habitat transition, including among lineages of Methylophilaceae (Betaproteobacteria) and a highly-studied Alphaproteobacteria (SAR11) moving to pelagic environments from soils, sediments, or other aquatic environments (39, 73, 77, 95). In these lineages, decrease in genome size is accompanied by the formation of gene paralogs and changes in GC content, coding density, coding sequence length/overlap, and intergenic spacer length, features collectively termed 'genome streamlining' (39). Multiple, independent genome reductions are also evident among Planctomycetes, which have transitioned from sediment/soil to pelagic freshwater habitats numerous times over their evolutionary history (3). There, intermediate genome sizes among more recently transitioned lineages (3) suggest that genome reduction is ongoing and may be achieved more slowly over time, though without some of the features commonly associated with genome streamlining.

Genome reduction has also been observed in multiple lineages that have colonized the mammalian gut/oral cavity from environmental sources, including the Elusimicrobia, the Melainabacteria, some chordate-associated Chlamydiae, and at least one archaeal methanogen (20, 28, 59, 90). Similarly, Saccharibacteria (CPR) adapted to the gut/oral environment show decreased genome sizes compared to relatives in soil and other environmental microbiomes (44), implying further reduction to already ancestrally small genomes (15) during the process of adaptation. Reduced metabolic repertoires in host-associated organisms may in part be a response to relatively nutrient-rich micro-environments, particularly the gut (90). This is an interesting contrast to the relatively oligotrophic pelagic habitats, where genome reduction may be driven by different factors, e.g. nutrient scarcity that limits microbial growth (61). Interestingly, adaptation to both the gut and pelagic water column are commonly linked to changes to environmental sensing and motility, including the loss of genes involved in chemotaxis and synthesis of the flagellar apparatus (Table 1). Additionally, some gut archaea have acquired membrane components that may facilitate attachment to host sites or other commensal bacteria (reviewed in 9).

| Transition type | Domain | Lineage(s) | Genome size/organization | Stress response | Metabolism | Motility/Attachment | References |
|---|---|---|---|---|---|---|---|
| environment -> oral/gut | Archaea | Various methanogens | | + bile salt hydrolases | - nitrogenase, CO oxidation, amino acid biosynthesis, some methanogenesis related genes<br>+ pyruvate/peptide uptake | + adhesin-like proteins and glycosyltransferases<br>- chemotaxis, motility genes | Borrel et al. 2020, Thomas et al. 2022 |
| environment -> oral/gut | Bacteria | Elusimicrobia | genome reduction | | - metabolic flexibility | | Meheust et al. 2020 |
| environment -> oral/gut | Bacteria | Melainabacteria | genome reduction | | - amino acid biosynthesis, nitrogen fixation, acetate production | - chemotaxis, flagellar apparatus | Di Rienzi et al. 2013 |
| environment -> oral/gut | Bacteria | Saccharibacteria (CPR) | genome reduction | + peroxide detoxification<br>- Fe-Mn superoxide dismutase, cytochrome o ubiquinol oxidase | + enolase, ribonucleoside-triphosphate reductase, lactate dehydrogenase, Cas genes, amino acid transport | | Jaffe et al. 2021, Mclean et al. 2020 |
| marine -> terrestrial | Bacteria | Flavobacteriaceae | | - $Na^+$-dependent transporters<br>+ $K^+$-dependent ATPase | + NDH complex, pectate lyase, urease utilization, glycoside hydrolase 25<br>- pyruvate carboxylase, glycoside hydrolase 16 | | Simon et al. 2017 |
| marine -> terrestrial | Bacteria | Rhodobacteraceae | | | - CO oxidation, B12 biosynthesis<br>+ sulfate uptake, formaldehyde oxidation, 6-phosphofructokinase | | Zhang et al. 2019 |
| marine -> freshwater | Bacteria | SAR11 (Alphaproteobacteria) | | | + EMP pathway<br>- ED pathway, glyoxylate shunt, other carbon degradation | | Eiler et al. 2016 |
| terrestrial -> marine | Bacteria | Salinispora (Actinobacteria) | | + sodium transporters<br>- large mechanosensitive channel | + electron transport components (NDH-1 complex) | | Penn & Jensen 2012 |
| sediment/soil -> freshwater/marine pelagic | Bacteria | Methylophilaceae (Betaproteobacteria) | genome reduction | + ectoine biosynthesis, sodium symporters (marine) | - nitrate reduction, cobalamin synthesis, methylotrophic capacities, formaldehyde oxidation<br>+ rhodopsin, NQR (marine) | - flagellar apparatus, chemotaxis | Salcher et al. 2019 |
| sediment/soil -> freshwater/marine pelagic | Bacteria | Planctomycetes | genome reduction, shorter intergenic regions, higher coding density | | + transporters for nitrogenous compound uptake, signal transduction | - cell/surface interaction genes, chemotaxis, flagellar apparatus | Andrei et al. 2019 |
| aerobic, high-temperature terrestrial -> moderate temperature, shallow marine | Archaea | Thaumarchaeota | | + $K^+$ transporter, ectoine biosynthesis/uptake, proline/glycine-betaine synthesis/uptake, oxidative/nitrosative stress components, small mechanosensitive channel<br>- reverse gyrase and other thermoprotectants, large mechanosensitive channel | | | Ren et al. 2019, Abby et al. 2020 |
| anaerobic terrestrial -> aerobic, high-temperature terrestrial | Archaea | Thaumarchaeota | | | + ammonia oxidation, cobalamin/biotin biosynthesis, urea utilization, amino acid biosynthesis<br>- nitrate/sulfate reduction, Wood-Ljungdahl pathway | | Ren et al. 2019, Abby et al. 2020, Sheridan et al. 2020 |

**Table 1.** Genomic features associated with habitat transitions in a subset of lineages. Features following a plus sign (+) indicate those features that are positively associated with the listed transition, i.e. enriched or gained in lineages adapting to the endpoint habitat. In contrast, features following a minus sign (-) indicate those that are negatively associated with a habitat transition, e.g. depleted or lost in the endpoint habitat.

## IV. The timing and tempo of habitat transition

While the nature of genetic changes - particularly, gene gains and losses - associated with habitat transitions is increasingly well understood, when these changes occur during lineage evolution is largely unknown. In principle, changes in gene content could precede, coincide with, or follow changes in habitat preference. In the first case, losses or acquisitions might occur on the phylogenetic branches immediately preceding habitat change, shifting metabolic capacities or stress responses that might enable successful establishment upon dispersal (Fig. 2a). Alternatively, changes to gene content might immediately post-date lineage divergence in a new habitat (Fig. 2b), potentially as a rapid phenomenon that enables survival, or via more gradual adaptation or genetic drift after a founding dispersal event (Fig. 2c). It is probable that lineages that acquire/lose genes prior to, or at the time of habitat transition also experience gain or loss events that postdate transitions as well (Fig. 2ab).

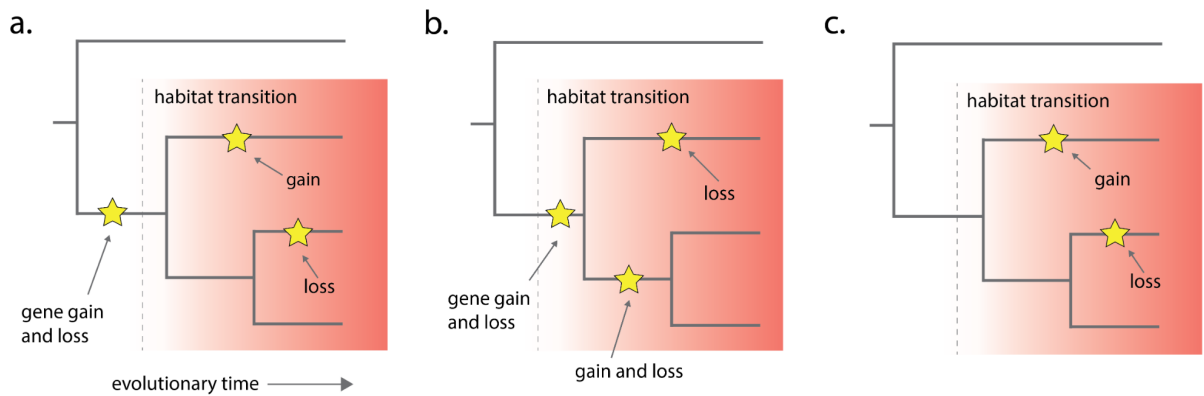

**Figure 2.** The relative timing of habitat transition and changes to gene inventory. Dashed lines in each panel indicate a successful habitat transition for the lineage, with evolutionary events (gene gains, losses, etc.) **a)** preceding the transition **b)** immediately postdating the transition and **c)** gradually postdating the transition.

More recently, a set of phylogenetic techniques called 'gene-species tree reconciliation' has begun to address these possibilities in a subset of lineages. These techniques compare the evolutionary histories of individual genes of interest to those of lineages more broadly (86), placing evolutionary events impacting those gene families (e.g., transfer, duplication, loss) onto the species tree. So far, most evidence suggests that habitat-specific, potentially adaptive traits are acquired after lineage transition into new habitats, often by lateral transfer from indigenous microbes already genomically adapted to the new environmental medium (44, 70, 77). These findings raise the possibility that lineages with a baseline degree of metabolic flexibility, stress tolerance, and ability to uptake exogenous DNA may be more successful at establishing within a new environment after dispersal. Alternatively, dispersers might simply survive and persist long enough to acquire adaptive traits by random chance alone (96).

As previously discussed, some habitat transitions may involve genomic shifts that are simply too complex to be made by new gene acquisition alone (55). More physiologically demanding transitions, e.g. across salinity or pH barriers, thus may require a more gradual series of coordinated changes over time, potentially beginning prior to transition itself. Supporting this possibility, Eiler et al. (2016) found that at least one major shift in carbon metabolism associated with the transition of SAR11 (Alphaproteobacteria) into freshwater had occurred in a common ancestor with marine progenitors. To date, though, very little other evidence argues for the role of early-occurring genomic changes in habitat change. This in part may be due to a methodological focus on trait presence/absence that may obscure subtle but important tweaks to existing gene sequences that allow for niche change (19), e.g. expansion of substrate range for key enzymes. In addition to 'niche-transcending' mutations of existing genes, gene duplication may also play an important role in enabling transition by modification of existing genetic material (79, 82).

Regardless of the timing of events, gene gain, duplication, loss, and other processes likely worked together to remodel lineages over time in response to major habitat change (31, 44). Such a gradual process may have involved 'innovation' steps where lineages accrued important preliminary genomic changes that paved the way for subsequent 'acclimatization' to new conditions (3).

# V. Symbiosis/dependence as a modulator of habitat transitions

Thus far, we have mainly considered characteristics of microorganisms themselves that determine their ability to switch habitats, and the processes by which this occurs. How do these constraints change when a microbe is dependent on one or more other microorganisms in some way? Inter-organismal dependencies, ranging from metabolic "handoffs" (2) to symbiosis (72) are common features of microbial communities. Thus, it stands to reason that these dependent relationships have also impacted the frequency and directionality of habitat transitions over evolutionary time.

**Case study: bacteria from the Candidate Phyla Radiation**

An interesting case study for habitat change within a framework of inter-species dependence is that of the Candidate Phyla Radiation (CPR) bacteria. These ultra-small celled bacteria with small genomes have been detected across Earth's biomes and were originally predicted to be symbionts of other organisms based on incomplete or missing biosynthetic pathways for lipids, amino acids, and nucleotides (46). Subsequent genomic and experimental work has expanded the range of potential metabolic interdependencies between CPR cells and their microbial hosts to include acid modulation (92) and the provision of metabolic cofactors (43). Increasingly emerging is a picture of CPR as a highly phylogenetically diverse lineage that has evolved a range of lifestyle strategies, including obligate parasitic or predatory interactions (e.g. (64)), detrital scavenging, and symbiosis for acquisition of basic molecular building blocks from either bacterial or archaeal hosts (15, 17, 48).

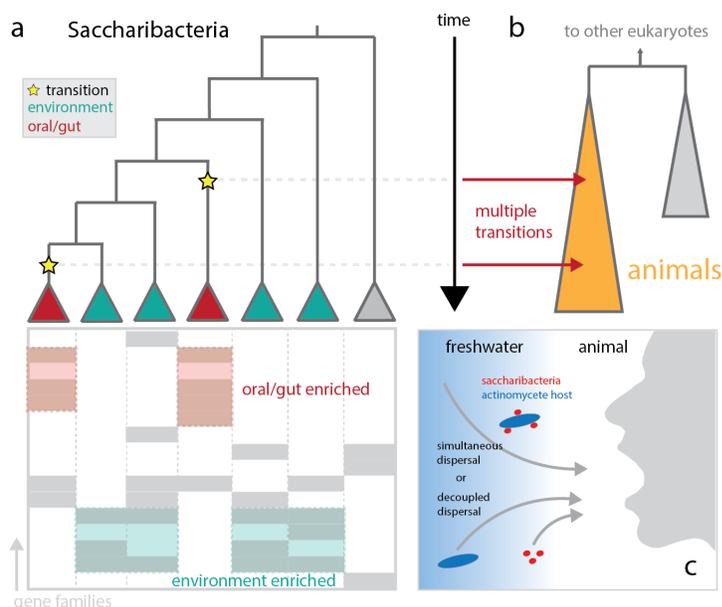

**Figure 3:** Habitat transition in the Saccharibacteria (Candidate Phyla Radiation). Phylogenetic evidence suggests that Saccharibacteria have transitioned from environmental microbiomes to animal mouths/guts multiple times over their evolutionary history (**ab**). Alongside genome reduction, transition to the animal microbiome likely involved the gain of some gene families and the loss of others relative to environmental counterparts (**a**). It is currently unknown whether Saccharibacteria colonized animals simultaneously with microbial hosts from freshwater, or did so independently and then acquired new hosts (**c**).

Drawing from the dozens of CPR lineages that have been recognized, we and others have previously detailed the evolutionary history of three - the Absconditabacteria, Gracilibacteria, and Saccharibacteria - that are commonly found in both environmental and host-associated microbiomes (58). Phylogenetic evidence suggests that these lineages have made independent migrations into the human/animal microbiome, and in two of the three cases, have done so multiple times (conceptualized in Fig. 3ab)(44, 58). The frequency of these transitions within CPR may be due to human and animal consumption of groundwater, where they are highly abundant (16, 23, 40). It is conceivable that transitions may have occurred via co-migration with existing host cells ('simultaneous dispersal'), or, alternatively, by acquisition of new, suitable host already adapted to the animal microbiome ('decoupled dispersal') (58, 81) (Fig. 3ab). While it is currently difficult to distinguish between these possibilities, several lines of evidence point towards new host acquisition as a favored model. First, some CPR-host associations appear to be flexible and phylogenetically diverse rather than highly conserved (22). Second, oral Saccharibacteria in co-culture with their Actinobacteria hosts can apparently "adapt" to new hosts after periods of host separation (8). Finally, species-level phylogenies for the Saccharibacteria and Actinobacteria show no strong signs of coevolution (44). Longer term, however, more detailed studies on the nature of CPR-host relationships are needed to fully understand factors driving animal colonization, especially given their emerging associations with human disease (reviewed in 65).

Regardless of their mode of dispersal, CPR bacteria experienced numerous genomic changes during transition to animal microbiomes, including decreased genome size and acquisition of new, possibly adaptive genes via lateral transfer (44, 58). These potentially adaptive gene gains can in part be identified by arraying 'families' of evolutionarily related sequences alongside an annotated phylogenetic tree and measuring their degree of enrichment within genomes from a given habitat (demonstrated conceptually for Saccharibacteria in Fig. 3a). Gene families that are statistically 'enriched' in Saccharibacteria from animal microbiomes are generally not found in relatives from the environment (Fig. 3a). On the other hand, other gene families common among environmental lineages are absent from animal associated

Saccharibacteria ('environment-enriched'), suggesting that they may have been lost during the process of transition. While some gene families are consistent across lineages that underwent independent transitions into animal microbiomes (possible convergent adaptation), some groups of gene families are lineage specific (demonstrated with a subset of published data in Fig. 4). This intriguing observation points to the existence of multiple distinct mechanisms of animal adaptation, possibly related to host type (e.g. animal vs. human) or body site (e.g. gut vs. oral), possibly due to origination from different lineages (Fig. 4).

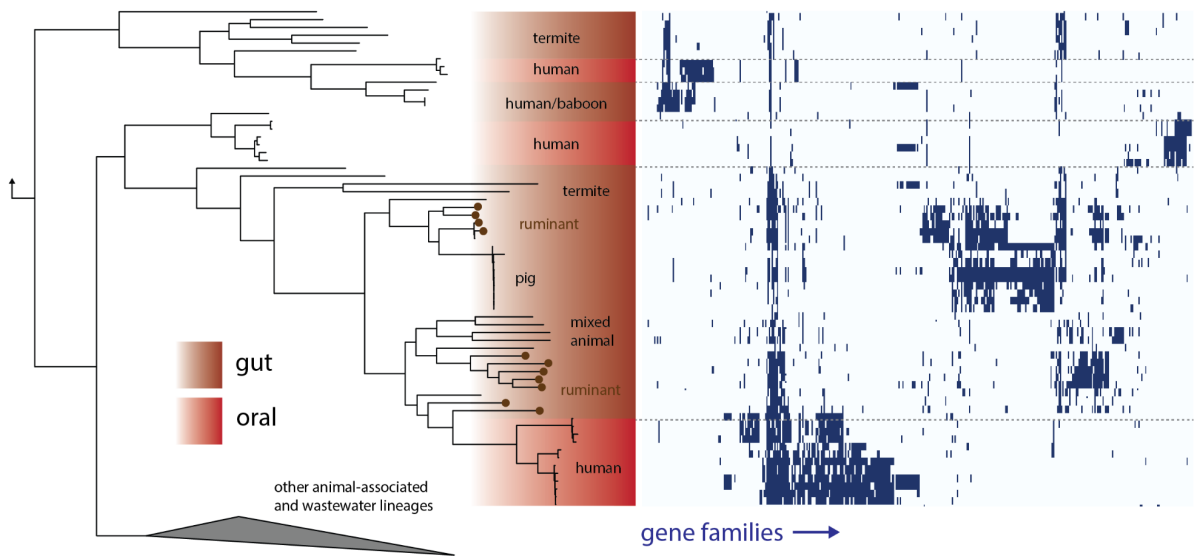

**Figure 4.** Lineage-specific gene families in Saccharibacteria (CPR) from different animal-associated microbiomes (adapted from Jaffe et al. 2021). At left is a subclade from a broader phylogenetic tree of Saccharibacteria that focuses on gut and oral lineages, annotated with their host type. Organisms from ruminant microbiomes are further indicated with brown bubbles. At right, a subset of gene families specific to subclades of Saccharibacteria are indicated, with dark shading indicating presence in a given genome, and light shading indicating absence.

**Other forms of microbial dependency**

One clear parallel to the case of CPR bacteria are prokaryotic viruses that depend on their bacterial and archaeal hosts for replication. Studies of viral biogeography show that while many prokaryotic viruses are habitat-specific, some are distributed across multiple habitat types (68). It remains unclear if broader distribution patterns simply reflect a broad distribution of hosts, or if the pattern is due to widened host range. Indeed, while relatively rare, some viruses can infect multiple hosts across higher-ranking phylogenetic groups, including across phyla (25, 68). Similar questions remain for intracellular Chlamydiae, which parasitize an incredible diversity of single-celled and multicellular eukaryotic hosts across ecosystems (20, 47). In general, we speculate that wide host range has been a key determinant of habitat transition across the spectrum of microbial dependence, including for CPR bacteria and phages.

One of the relatively few groups of phages that has been studied in the framework of habitat transition are the Lak megaphages, which replicate in *Prevotella*, a bacterial group that is prevalent in the guts of humans that consume a non-Western diet and in some animals (26). An analysis of the microbiomes of diverse animals and various human cohorts showed that Lak phages have migrated into the gut environments of pigs, dogs, horses, many different primates, tortoises, and humans numerous times over the course of animal evolution (26). Notably, the predicted proteomes of the Lak phages are animal specific, suggesting acquisition of unique gene complements alongside animal host switches (regardless of whether this occurred dependently or independently of *Prevotella* hosts) (21).

## Outlook and future work

Habitat transitions are common occurrences in the microbial world with important ramifications for lineage evolution and host/ecosystem health. However, to date, such transitions remain unstudied or understudied in most bacterial, archaeal, and viral lineages. Thus, expanding the slate of groups subject to comparative studies across environments should be a first priority for future research. Fortunately, the tidal wave of genome information now accumulating from virtually every possible ecosystem makes the power of such analyses ever greater. Study of additional lineages, particularly those informed by gene/protein family analyses, will undoubtedly deepen our understanding of emerging commonalities (e.g., change in genome size, loss of metabolic versatility) among varied microbial habitat transitions, including specific genomic adaptations. However, we expect that such work will also likely reveal intriguing idiosyncrasies that broaden our understanding of the 'rules of life', particularly along the spectrum of microbial dependence (Section V above).

Among the many remaining knowledge gaps concerning microbial habitat transitions, one of the greatest concerns the origin and subsequent evolution of animal-associated lineages. In recent years, the structure and diversity of the animal associated microbiome has been intensely studied due to its emerging links to a wide range of human health conditions (32). Despite a growing understanding of how the human microbiome assembles and subsequently changes on the order of a single human lifetime, relatively little is known about the deep evolutionary origins of bacterial and archaeal taxa that evolved to be commensals in various animal guts, mouths, and skin. How did environmental microbes originally colonize and establish in these microbiomes over the course of animal evolution? When did these events occur? Addressing this question will not only shed light on important past events in the establishment of the human microbiota, but should also carry relevance for our understanding of future disease, the spread of antibiotic resistance, and the potential efficacy and persistence of probiotics.

While our discussion here has largely focused on ancient events that led to the spread of microbial lineages across planetary microbiomes, we acknowledge that habitat transition is likely ongoing and occurring at multiple timescales, including those transitions into the human microbiome from environmental sources. Together with host genetics and diet, early microbial colonizers to the human gut - including those acquired from the surrounding environment - play significant roles in subsequent microbiome assembly and human development (36, 78). For example, using stringent genome-to-genome comparisons, it has been demonstrated that hospital-associated microbial strains move between the built environment of the neonatal intensive care unit and sequential premature infant room occupants (11, 12). These horizontal transmissions may constitute independent habitat transitions that occur on short timescales, although environmentally-acquired strains are often less likely to persist into later life than vertically acquired, parental strains (34, 54). Similarly, human pathogens derived from environmental reservoirs - e.g. the colonization of the lungs by *Pseudomonas aeruginosa* or *Burkholderia dolosa* in cystic fibrosis patients (33, 50), or the colonization of gastrointestinal tract by *Burkholderia pseudomallei* from soil and water sources (5) - also undergo a form of habitat transition during host infection. Given the constant influx of environmental lineages to the developing animal microbiome over time, it is interesting to consider how frequently and by what mechanisms organisms have evolved towards different points on the spectrum between mutualism and pathogenicity.

Under certain circumstances, it is also conceivable that host-adapted microbes may occasionally transition back to environmental reservoirs from which they originally evolved. Possible routes for such transmission include the seeding of soil or wastewater microbiomes by animal waste, or the dispersal of host-associated microbes into the surrounding medium following host death/decomposition (71). However, it is currently unknown whether lineages experiencing such dispersal events are able to permanently establish in the recipient biome or are merely transient passers-by. Future work targeted on possible hotspots of such transitions could establish the frequency/longevity of such events and determine whether they can give rise to new lineages with different genomic characteristics than their ancestors.

## Acknowledgments

We thank our colleagues Yue Clare Lou, Alexander Crits-Christoph, and Jett Liu for helpful discussions and their feedback on the manuscript, and Guillaume Borrel, Raphaël Méheust, and Michaela Salcher for their aid in reproducing the phylogenetic trees included in Figure 1. We also thank Leslie Parker and the editorial staff of Annual Review of Microbiology for their guidance throughout the writing process. A.L.J. was supported by the Stanford Science Fellows program and the NSF Postdoctoral Research Fellowship in Ocean Sciences. We acknowledge additional support from the Moore Foundation program on aquatic symbiosis, Chan Zuckerberg Biohub, and the National Institutes of Health.